\documentclass[fleqn,usenatbib]{mnras}

\usepackage{newtxtext,newtxmath,xcolor}

\usepackage[T1]{fontenc}
\usepackage{ae,aecompl}

\usepackage{graphicx}	
\usepackage{amsmath}	
\usepackage{amssymb}	

\def \sb {mag\,arcsec$^{-2}$}
\def \mue {$\langle\mu\rangle_{e}$}
\def \ellip {$\overline{E}$}
\def \triax {$\overline{T}$}
\def \sigellip {$\sigma_{E}$}
\def \sigtriax {$\sigma_{T}$}
\def \galfit {{\sc galfit}}

\def \emcee {{\sc emcee}}
\def \msun {M$_{\odot}$}
\def \lsun {$L_{\odot}$}

\def \mstar {M$_{*}$}

\newcommand{\Evirgo}{0.43}
\newcommand{\Euvirgo}{+0.02}
\newcommand{\Elvirgo}{-0.02}
\newcommand{\SEvirgo}{0.12}
\newcommand{\SEuvirgo}{+0.02}
\newcommand{\SElvirgo}{-0.01}
\newcommand{\Tvirgo}{0.15}
\newcommand{\Tuvirgo}{+0.05}
\newcommand{\Tlvirgo}{-0.05}
\newcommand{\STvirgo}{0.04}
\newcommand{\STuvirgo}{+0.10}
\newcommand{\STlvirgo}{-0.02}

\newcommand{\Efornax}{0.46}
\newcommand{\Eufornax}{+0.02}
\newcommand{\Elfornax}{-0.03}
\newcommand{\SEfornax}{0.13}
\newcommand{\SEufornax}{+0.03}
\newcommand{\SElfornax}{-0.02}
\newcommand{\Tfornax}{0.24}
\newcommand{\Tufornax}{+0.46}
\newcommand{\Tlfornax}{-0.14}
\newcommand{\STfornax}{0.13}
\newcommand{\STufornax}{+0.19}
\newcommand{\STlfornax}{-0.10}

\newcommand{\Ecena}{0.43}
\newcommand{\Eucena}{+0.08}
\newcommand{\Elcena}{-0.07}
\newcommand{\SEcena}{0.15}
\newcommand{\SEucena}{+0.08}
\newcommand{\SElcena}{-0.07}
\newcommand{\Tcena}{0.26}
\newcommand{\Tucena}{+0.46}
\newcommand{\Tlcena}{-0.16}
\newcommand{\STcena}{0.07}
\newcommand{\STucena}{+0.19}
\newcommand{\STlcena}{-0.05}

\newcommand{\Elg}{0.50}
\newcommand{\Eulg}{+0.10}
\newcommand{\Ellg}{-0.08}
\newcommand{\SElg}{0.14}
\newcommand{\SEulg}{+0.07}
\newcommand{\SEllg}{-0.06}
\newcommand{\Tlg}{0.43}
\newcommand{\Tulg}{+0.26}
\newcommand{\Tllg}{-0.24}
\newcommand{\STlg}{0.09}
\newcommand{\STulg}{+0.20}
\newcommand{\STllg}{-0.07}

\newcommand{\EbN}{0.42}
\newcommand{\EubN}{+0.03}
\newcommand{\ElbN}{-0.03}
\newcommand{\SEbN}{0.14}
\newcommand{\SEubN}{+0.04}
\newcommand{\SElbN}{-0.03}
\newcommand{\TbN}{0.11}
\newcommand{\TubN}{+0.10}
\newcommand{\TlbN}{-0.07}
\newcommand{\STbN}{0.05}
\newcommand{\STubN}{+0.11}
\newcommand{\STlbN}{-0.03}

\newcommand{\EbnN}{0.50}
\newcommand{\EubnN}{+0.03}
\newcommand{\ElbnN}{-0.03}
\newcommand{\SEbnN}{0.12}
\newcommand{\SEubnN}{+0.03}
\newcommand{\SElbnN}{-0.03}
\newcommand{\TbnN}{0.57}
\newcommand{\TubnN}{+0.18}
\newcommand{\TlbnN}{-0.15}
\newcommand{\STbnN}{0.09}
\newcommand{\STubnN}{+0.22}
\newcommand{\STlbnN}{-0.07}

\newcommand{\EfN}{0.37}
\newcommand{\EufN}{+0.03}
\newcommand{\ElfN}{-0.03}
\newcommand{\SEfN}{0.11}
\newcommand{\SEufN}{+0.04}
\newcommand{\SElfN}{-0.03}
\newcommand{\TfN}{0.22}
\newcommand{\TufN}{+0.37}
\newcommand{\TlfN}{-0.12}
\newcommand{\STfN}{0.08}
\newcommand{\STufN}{+0.21}
\newcommand{\STlfN}{-0.06}

\newcommand{\EfnN}{0.45}
\newcommand{\EufnN}{+0.02}
\newcommand{\ElfnN}{-0.02}
\newcommand{\SEfnN}{0.12}
\newcommand{\SEufnN}{+0.03}
\newcommand{\SElfnN}{-0.02}
\newcommand{\TfnN}{0.23}
\newcommand{\TufnN}{+0.20}
\newcommand{\TlfnN}{-0.11}
\newcommand{\STfnN}{0.14}
\newcommand{\STufnN}{+0.19}
\newcommand{\STlfnN}{-0.11}

\defcitealias{Lisker2007}{L07}

\title[The shapes of faint dwarf galaxies]{How nucleation and luminosity shape faint dwarf galaxies}

\author[S\'anchez-Janssen et al.]{R.\ S\'anchez-Janssen,$^{1}$\thanks{E-mail: ruben.sanchez-janssen@stfc.ac.uk}
T.~H.\ Puzia,$^{2}$
L.\ Ferrarese,$^{3}$
P.\ C\^ot\'e,$^{3}$
P.\ Eigenthaler,$^{2}$\newauthor
B.\ Miller,$^{4}$
Y.\ Ordenes-Brice\~no,$^{2}$
E.~W.\ Peng,$^{5}$
K.~X.\ Ribbeck,$^{2}$
J.\ Roediger,$^{3}$\newauthor
C.\ Spengler,$^{2,5,6}$
M.~A.\ Taylor.$^{7}$
\\
$^{1}$UK Astronomy Technology Centre, Royal Observatory, Blackford Hill, Edinburgh, EH9 3HJ, UK\\
$^{2}$Institute of Astrophysics, Pontificia Universidad Cat\'olica de Chile, Av.~Vicu\~na Mackenna 4860, 7820436 Macul, Santiago, Chile\\
$^{3}$NRC-Herzberg Astronomy and Astrophysics, 5071 West Saanich Road, Victoria, BC, V9E 2E7, Canada\\
$^{4}$Gemini Observatory, Southern Operations Center, Casilla 603, La Serena, Chile\\
$^{5}$Department of Astronomy and Kavli Institute for Astronomy and Astrophysics, Peking University, Beijing 100871, China\\
$^{6}$Department of Physics and Astronomy, University of Victoria, Victoria, BC V8P 5C2, Canada\\
$^{7}$Gemini Observatory, Northern Operations Center, 670 North A'ohoku Place, Hilo, HI 96720, USA
}

\date{Accepted 2019 January 11. Received 2019 January 10; in original form 2018 November 1}

\pubyear{\the\year}

\begin{document}
\label{firstpage}
\pagerange{\pageref{firstpage}--\pageref{lastpage}}
\maketitle

\begin{abstract}
We study the intrinsic shapes of a sample of over 400 quiescent galaxies in the cores of the Virgo and Fornax clusters with luminosities $10^{6} \leq L_{g}/L_{\odot} \leq 10^{8}$.
Similar to satellites of the Local Group and Centaurus\,A, these faint, low surface brightness cluster galaxies are best described as a family of thick ($\langle C/A \rangle > 0.5$), oblate-triaxial spheroids.
However, the large sample size allows us to show  that the flattening of their stellar distribution depends both on luminosity and on the presence of a nuclear star cluster. 
Nucleated satellites are thicker at all luminosities compared to their non-nucleated counterparts, and fainter galaxies are systematically thicker as well, regardless of nucleation. 
Once nucleation is accounted for, we find no evidence that the environment the satellites live in plays a relevant role in setting their three-dimensional structure.
We interpret both the presence of stellar nuclei and the associated thicker shapes as the result of preferential early and rapid formation, effectively making these faint nucleated galaxies the first generation of cluster satellites. 


\end{abstract}




\section{Introduction}
The three-dimensional (3D) shapes of galaxies provide constraints on the different physical mechanisms that play a role in the process of galaxy formation and assembly.
Recent theoretical and numerical work indicates that the emergence of thin galactic discs is controlled by the gas fraction and the feedback from young stars, which in turn regulate star formation and chemical enrichment \citep{Hayward2015,Navarro2017,Ma2017}. 
As a result of their burstier star formation histories, cold rotating discs assemble later in lower mass galaxies \citep{Simons2017}.
At higher galaxy masses the contribution from dissipationless mergers grows significantly \citep{Oser2010,Clauwens2017}, and the resulting deposition of accreted stars leads to the emergence of spheroids, which notably exhibit a high degree of asphericity at the massive end \citep{Holden2012,Foster2017,Krajnovic2018,Li2018}.
Major mergers are not relevant for the assembly of very low-mass galaxies \citep{Fitts2018}, but it is well established now that the stellar distributions of isolated \mstar\ $\lesssim 10^{9}$ \msun\ galaxies are significantly thicker than what is typical for rotating cold discs \citep{Sanchez-Janssen2010a,Roychowdhury2013}.
This suggests that in these systems dispersion support is dominant over rotation \citep{Wheeler2015}, and points towards feedback mechanisms setting the internal structure of  low-mass centrals \citep{Kaufmann2007}. 
Explosive feedback episodes from supernovae and intense star formation act to dynamically heat preexisting stellar populations, thus driving stellar migration and thickening of the galactic bodies \citep{Governato2010,Pontzen2012,El-Badry2015}. 
In the case of satellite dwarfs additional hydrodynamical and tidal interactions  with their hosts can  modify the structural and dynamical properties of low-mass systems, creating pressure-supported triaxial objects that in many aspects resemble the  satellite populations of groups and clusters (\citealt{Mastropietro2005,Smith2010,Kazantzidis2011,Kazantzidis2017,Lokas2012}; but see \citealt{Smith2015}).

Until recently the 3D structure of faint satellites have been poorly studied. 
This is due primarily to their low luminosities and surface brightness levels, which limit the detail with which they can be studied.
Because the orientation of individual galaxies in the sky is almost always unknown, the precise intrinsic shape of a given object can only be inferred by a combination of structural and kinematical data \citep{Franx1991}.
The latter is extremely hard to obtain for faint systems, for which  shapes are usually derived on statistical grounds via inversion or modelling of the apparent ellipticity distribution for an entire population \citep[][L07 hereafter]{Sandage1970,Ryden1994,Lisker2007}.
Naturally, the Local Group (LG) satellites are the best characterised population of low-mass galaxies. 
'Classical' dwarf spheroidals ($M_{V} \lesssim -8.5$) are a class of relatively thick galaxies with intrinsic flattening $\sim$\,0.5 and a certain degree of triaxiality \citep{Salomon2015,Sanchez-Janssen2016a,Sanders2017}.
The advent of wide, deep imaging surveys has finally allowed mapping of larger volumes down to sufficiently low luminosities to enable comparable studies in populations of dwarf galaxies outside the LG \citep{Ferrarese2012,Taylor2017,Danieli2017,Eigenthaler2018,Venhola2018}.
These datasets not only offer us the opportunity to study faint galaxies in other environments, but also with much larger number statistics than is possible locally--a critical aspect if one is to investigate how their properties depend on mass or galaxy type.

Beyond the simple distinction between star-forming and quiescent dwarfs, the latter class are known to be quite a diverse group of galaxies, with some subpopulations featuring the presence of nuclear star clusters (NSCs) or even disc-like features \citepalias{Lisker2007}.
NSCs are compact stellar systems that in many aspects resemble typical globular clusters (GCs), but tend to be larger, more massive, and can host complex stellar populations, even in low-mass quiescent galaxies \citep{Cote2006,Lisker2007,Turner2012,Paudel2011,Spengler2017}. 
NSCs are found in galaxies of all morphological types, but tend to be particularly common in high-density environments.
Indeed, in galaxy clusters like Virgo and Fornax, nucleation persists down to extremely low galaxy luminosities, $L_{V} \approx 10^{6}$ \lsun\ \citep{Sanchez-Janssen2018,Ordenes-Briceno2018b}.
Regardless of their exact formation mechanism, the presence of NSCs is indicative of past high-pressure star formation episodes, and thus provide clues on the formation of their host galaxy.
Nucleated quiescent galaxies with stellar masses log(\mstar/\msun) $\!\approx\!$ 8-9 have been shown to exhibit distinct clustering and phase-space properties \citep{Ferguson1989} as well as to host the oldest stellar populations among all satellite subclasses \citep{Schombert2006,Lisker2008}.
Non-nucleated and nucleated intermediate-mass cluster satellites are best described as families of thick ($\langle C/A\rangle \approx 0.5$), nearly oblate spheroids, but the former subpopulation is known to be substantially flatter than the latter \citep{Ryden1994}. 
\citetalias{Lisker2007} find that, at least for non-nucleated Virgo dEs, mass also seems to play a role, with fainter satellites being thicker than their more luminous counterparts.

In this contribution we take advantage of the present wealth of data on faint satellites outside the LG to show that these trends hold down to $L_{V}\! \approx\! 10^{6}$ \lsun. The combined sample size is large enough that it allows us, for the first time in this luminosity regime, to investigate shape differences as a function of brightness and nucleation.

\section{Satellites in the Local Universe}

The galaxy samples used in this investigation come from the Next Generation Virgo Cluster Survey \citep[][NGVS]{Ferrarese2012}, the Next Generation Fornax Survey \citep[][NGFS]{Eigenthaler2018}, and the Survey of Centaurus A's Baryonic Substructures \citep[][SCABS]{Taylor2017}.
These studies have imaged these elliptical-dominated environments out to their respective virial radii, uncovering rich populations of mostly quiescent low-mass galaxies.
Very detailed descriptions of the surveys are provided elsewhere, and here we only mention briefly the sample selection, the derivation of structural parameters, and the detection of nuclei.

The Virgo sample is identical to that presented in \citet{Sanchez-Janssen2016a}. 
Galaxies in the central $R \lesssim 0.2R_{\rm vir}$ of the cluster are detected on $ugriz$ images using a ring-median filter, and their two-dimensional light distributions are modelled as \citet{Sersic1968} functions with \galfit\ \citep{Peng2002}.
If the fit is not satisfactory due to the presence of an NSC, a second PSF-like component is included in the model.  
The identification of nucleated satellites is further complemented by a visual inspection of colour and unsharp-masked images of the galaxies \citep[][]{Sanchez-Janssen2018}. 
We construct our sample of quiescent Virgo satellites from galaxies that have colours consistent with the cluster red sequence \citep{Roediger2016}.

\citet{Eigenthaler2018} present the properties of the dwarf galaxy population in the central $R \lesssim 0.25R_{\rm vir}$ of Fornax.
Galaxy detection is carried out by visual inspection of stacked $ugi$ colour images, and candidates are selected according to their diffuse morphologies and for not showing evidence of ongoing star formation. 
Galaxies are also modelled with single-S\'ersic components, and the fitting procedure is carried out iteratively using \galfit. 
NSCs are identified visually as well, and in these cases a central mask is applied at each step of the fitting process to account for their presence.  

The SCABS data cover an area $R \lesssim 0.5R_{\rm vir}$ around NGC\,5128/CenA \citep[][]{Taylor2017,Taylor2018}. The dwarfs were selected and their morphological parameters derived in an identical way to the NGFS (Ribbeck et al.~2019).

Finally, we use LG satellites as representatives of a lower-density environment. 
The parent sample comes from \citet{McConnachie2012}, but only systems within the zero-velocity surface of the LG are selected (see their Fig.\,5).
In addition, we only keep galaxies with low gas-mass fraction, M$_{\text{H{\sc i}}}/L_{V} \leq 0.1~\text{M}_{\odot}/L_{\odot}$, which mimics our selection of quiescent cluster galaxies.
The axis ratio measurements and their uncertainties are taken directly  from \citet{McConnachie2012}.

We emphasise that the detection of NSCs in all samples is very robust. NSCs have much higher surface brightness than the underlying galaxy light,  with a median $\Delta\mu$ = 2.4mag and $\sigma_{\Delta\mu}$ = 1.3mag. 
The distribution of $\Delta\mu$ shows no dependence on $q$, and we detect NSCs across the entire range of observed axis ratios (see Fig.\,\ref{fig:ba}).
Throughout this work we only use magnitudes and axis ratios measured in the $g$-band.
Given that we use the same fitting function and code to model satellites in Virgo, Fornax and CenA, we also apply the same \mue-dependent uncertainty for the axis ratio measurements of the two populations. 
The amplitude of these uncertainties is obtained from  \galfit\ modelling of simulated single-S\'ersic galaxies, and it ranges from a minimum value of $\sigma_{q} =0.02$ to 0.15 at \mue\ = 27.75\,\sb\ (the NGVS 50\% completeness limit).
For all  samples we impose the same criterion for low luminosity, $-15 \leq M_{g} \leq -10$, where we assume  $(g-V)=0.2$ for all LG satellites. 
The number of galaxies in each environment is shown in the second column of Table\,\ref{tab:results}.
We note that the Sagittarius dSph is the only LG satellite that satisfies all the above conditions and can be considered as nucleated.
Only one of the CenA satellites hosts a nucleus, and in the following we treat all the group satellites essentially as a non-nucleated sample.

\section{Axis ratio distributions}

\begin{figure}
\includegraphics[width=\columnwidth]{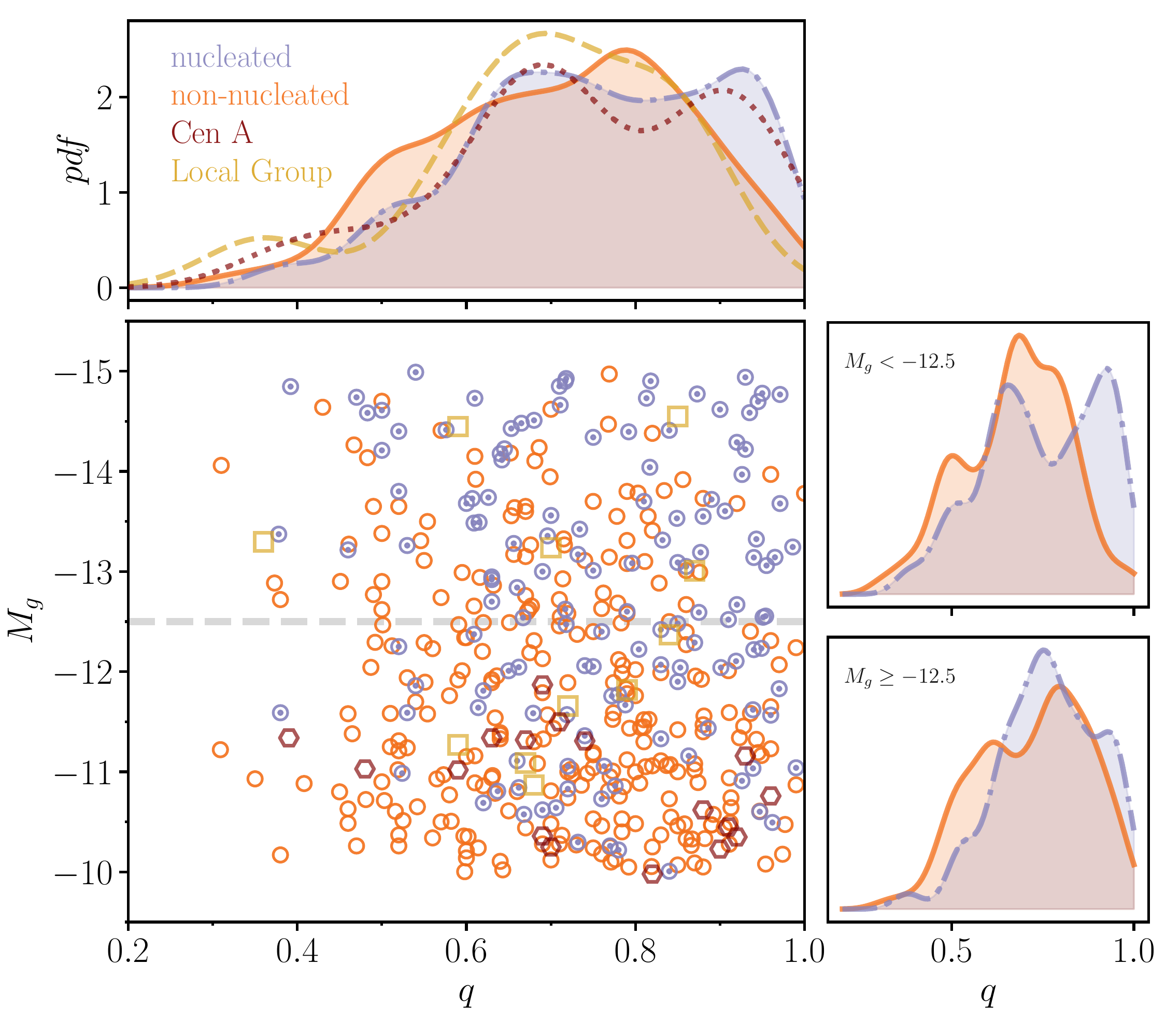}
\caption{
{\it Central panel:} Distribution of apparent axis ratios ($q$) and absolute magnitude ($M_{g}$) for dwarf galaxies in Virgo and Fornax (circles), CenA (hexagons) and the LG (squares). Open circles correspond to non-nucleated cluster dwarfs, whereas the galaxies hosting NSCs contain an inner dot symbol. The dashed horizontal line at $M_{g} = -12.5$ shows the division between the samples of bright and faint dwarfs. Notice the luminosity dependence of nucleation.
{\it Top:} Probability distribution of apparent axis ratios for the four different populations of dwarfs. The paucity of apparently round non-nucleated dwarfs compared to nucleated ones is evident. 
{\it Right:} These two panels show, for dwarfs in Virgo and Fornax only, the axis ratio pdf for bright (top) and faint (bottom) non-nucleated (solid) and nucleated dwarfs (dot-dashed). 
}
\label{fig:ba}
\end{figure}

The central panel in Figure\,\ref{fig:ba} shows the distribution of absolute magnitudes in the $g$-band and apparent axis ratios for dwarfs in Virgo and Fornax. 
We do not distinguish between the three populations, but focus on whether the dwarf exhibits an NSC (dot-circle symbols) or not (open circles). 
This is justified by our finding in Section\,\ref{sect:shapes} that the satellites in all environments have consistent intrinsic shapes.
For reference, we also plot quiescent satellites in the LG (squares) and in CenA (hexagons). 
Two things are immediately obvious from this panel. 
First, and as already shown by \citet{Sanchez-Janssen2018} for Virgo and \citet{Ordenes-Briceno2018b} for Fornax, nucleation is a strong function of galaxy luminosity, with non-nucleated dwarfs being dominant towards fainter magnitudes.
Second, the distributions of apparent axis ratios differ for nucleated and non-nucleated dwarfs. 
This is perhaps best appreciated in the top panel of Fig.\,\ref{fig:ba}, where we show the probability distribution function (pdf) for the nucleated (dot-dashed line) and the non-nucleated (solid) cluster satellites, and the non-nucleated group satellites (dashed and dotted lines for the LG and CenA, respectively). 
In all cases the pdf is a simple Gaussian kernel density estimation (kde) of the $q$ distribution.
The pdf for nucleated objects shows a peak for very round apparent axis ratios ($q\gtrsim0.9$) that is absent  from the corresponding distributions for non-nucleated cluster dwarfs and the LG objects. 
This is known to correspond to 3D shapes with low triaxiality. 

The two panels at the right of Fig.\,\ref{fig:ba} explore the luminosity dependence of the $q$ distributions. 
The top panel corresponds to the bright subsample ($-15 \leq M_{g} < -12.5$), where the difference between nucleated and non-nucleated dwarfs is maximised. 
The pdf for the bright nucleated subsample is double-peaked, with maxima at $q\simeq0.65$ and $q\simeq0.95$.
While the lower peak is also present in the pdf of non-nucleated dwarfs, their distribution falls rapidly towards high-$q$ values.
The behaviour of these two distributions is mimicked by their faint counterparts ($-12.5 \leq M_{g} < -10$; bottom panel), but the difference between the subsamples is less significant. 
Not only do the main peaks of the distributions shift towards higher $q$ values, but the population of very round nucleated dwarfs is of relative lower importance.

\section{Intrinsic shapes inference}
\label{sect:shapes}

The analysis in the previous section provides a qualitative description of apparent shapes of nucleated and non-nucleated dwarfs at the very faint end of the galaxy luminosity function in nearby clusters and groups.
We now attempt to model the observed axis ratio distributions in order to constrain their 3D shapes. 
We follow the methodology presented in \citet{Sanchez-Janssen2016a}, whereby galaxies are modelled as a family of optically thin triaxial ellipsoids. 
We assume that the 3D galaxy density is structured as a set of coaligned ellipsoids characterised by a common ellipticity $E = 1 - C/A$, and a triaxiality $T = (A^{2} - B^{2})/(A^{2}-C^{2})$, where $A \ge B \ge C$ are the intrinsic major, intermediate, and minor axes of the ellipsoid, respectively \citep{Franx1991}.
The ellipticity and triaxiality parameters for the galaxy populations are assumed to be normally distributed, with means and standard deviations \ellip, \sigellip, \triax, and \sigtriax. 
We implement a Bayesian framework to explore the posterior distribution of the model parameters, which allows us to work directly on discrete data, and to account for individual, surface brightness-dependent axis ratio uncertainties.
We use flat priors in the interval $[0,1]$ for the location parameters of our model normal distributions (\ellip\ and \triax).  For the standard deviations \sigellip\  and \sigtriax\ we adopt scale-invariant priors of the form $p(\sigma) \propto \sigma^{-1}$.
We use the \emcee\ code \citep{ForemanMackey2013} to sample the posterior distribution of the parameters with 100 'walkers' and 1000 steps--sufficient for the MCMC chains to reach equilibrium. 

The results of this modelling are summarised in Table\,\ref{tab:results}, where we indicate the number of galaxies contributing to each subsample. 
The parameters for the preferred model correspond to the median and 68\% confidence values of the marginalised posteriors. 
The first result is that satellites in Virgo and Fornax are very similar populations in terms of their 3D stellar distributions. 
The parameters in Table\,\ref{tab:results} correspond to intrinsic axis ratios of 1:0.95:0.57 and 1:0.91:0.54, respectively.
It is well known that triaxiality distributions are not well constrained with photometric data alone \citep{Binggeli1980}, and we will therefore focus the bulk of our analysis on the comparison of galaxy flattenings (\ellip).
Nevertheless, it is worth mentioning that oblate-triaxial shapes are favoured in all cases over more prolate spheroids.

\begin{table}
    \centering
    \caption{Intrinsic shapes of faint quiescent satellites. We show results for samples in Virgo, Fornax, CenA, and the LG. For the cluster populations we further differentiate between nucleated (N) and non-nucleated (nN) bright (b) and faint (f) satellites.}
    \label{tab:results}
    \begin{tabular}{lcrrrr} 
        \hline
        Sample & \# & \ellip & \sigellip & \triax & \sigtriax\\
        \hline\hline
        Virgo & 228 & $\Evirgo^{\Euvirgo}_{\Elvirgo}$ & $\SEvirgo^{\SEuvirgo}_{\SElvirgo}$ & $\Tvirgo^{\Tuvirgo}_{\Tlvirgo}$ & $\STvirgo^{\STuvirgo}_{\STlvirgo}$\\
        Fornax & 185 & $\Efornax^{\Eufornax}_{\Elfornax}$ & $\SEfornax^{\SEufornax}_{\SElfornax}$ & $\Tfornax^{\Tufornax}_{\Tlfornax}$ & $\STfornax^{\STufornax}_{\STlfornax}$\\
        CenA & 16 & $\Ecena^{\Eucena}_{\Elcena}$ & $\SEcena^{\SEucena}_{\SElcena}$ & $\Tcena^{\Tucena}_{\Tlcena}$ & $\STcena^{\STucena}_{\STlcena}$\\
        LG & 11 & $\Elg^{\Eulg}_{\Ellg}$ & $\SElg^{\SEulg}_{\SEllg}$ & $\Tlg^{\Tulg}_{\Tllg}$ & $\STlg^{\STulg}_{\STllg}$\\
        \hline
        bN & 79 & $\EbN^{\EubN}_{\ElbN}$ & $\SEbN^{\SEubN}_{\SElbN}$ & $\TbN^{\TubN}_{\TlbN}$ & $\STbN^{\STubN}_{\STlbN}$\\
        bnN & 76 & $\EbnN^{\EubnN}_{\ElbnN}$ & $\SEbnN^{\SEubnN}_{\SElbnN}$ & $\TbnN^{\TubnN}_{\TlbnN}$ & $\STbnN^{\STubnN}_{\STlbnN}$\\
		fN & 58 & $\EfN^{\EufN}_{\ElfN}$ & $\SEfN^{\SEufN}_{\SElfN}$ & $\TfN^{\TufN}_{\TlfN}$ & $\STfN^{\STufN}_{\STlfN}$\\
        fnN & 200 & $\EfnN^{\EufnN}_{\ElfnN}$ & $\SEfnN^{\SEufnN}_{\SElfnN}$ & $\TfnN^{\TufnN}_{\TlfnN}$ & $\STfnN^{\STufnN}_{\STlfnN}$\\
        \hline
    \end{tabular}
\end{table}

This structural similarity allows us to combine the Virgo and Fornax cluster samples to investigate the dependence of intrinsic shapes on galaxy luminosity and nucleation. 
As indicated in Fig.\,\ref{fig:ba} and Table\,\ref{tab:results} we build four subsamples, with $M_{g} = -12.5$ serving as a dividing point between the bright (b) and faint (f) nucleated (N) and non-nucleated (nN) subpopulations. 
Despite the small subsample sizes the distributions of intrinsic ellipticity are remarkably well constrained in the range of \ellip~$\simeq$~0.35-0.5. 
When analysing the nucleated and non-nucleated cluster populations separately there is a trend for the latter to have a higher degree of triaxiality than the former.
This is consistent with the relative paucity of apparently round non-nucleated dwarfs (Fig.\,\ref{fig:ba}). 

\begin{figure}
\includegraphics[width=\columnwidth]{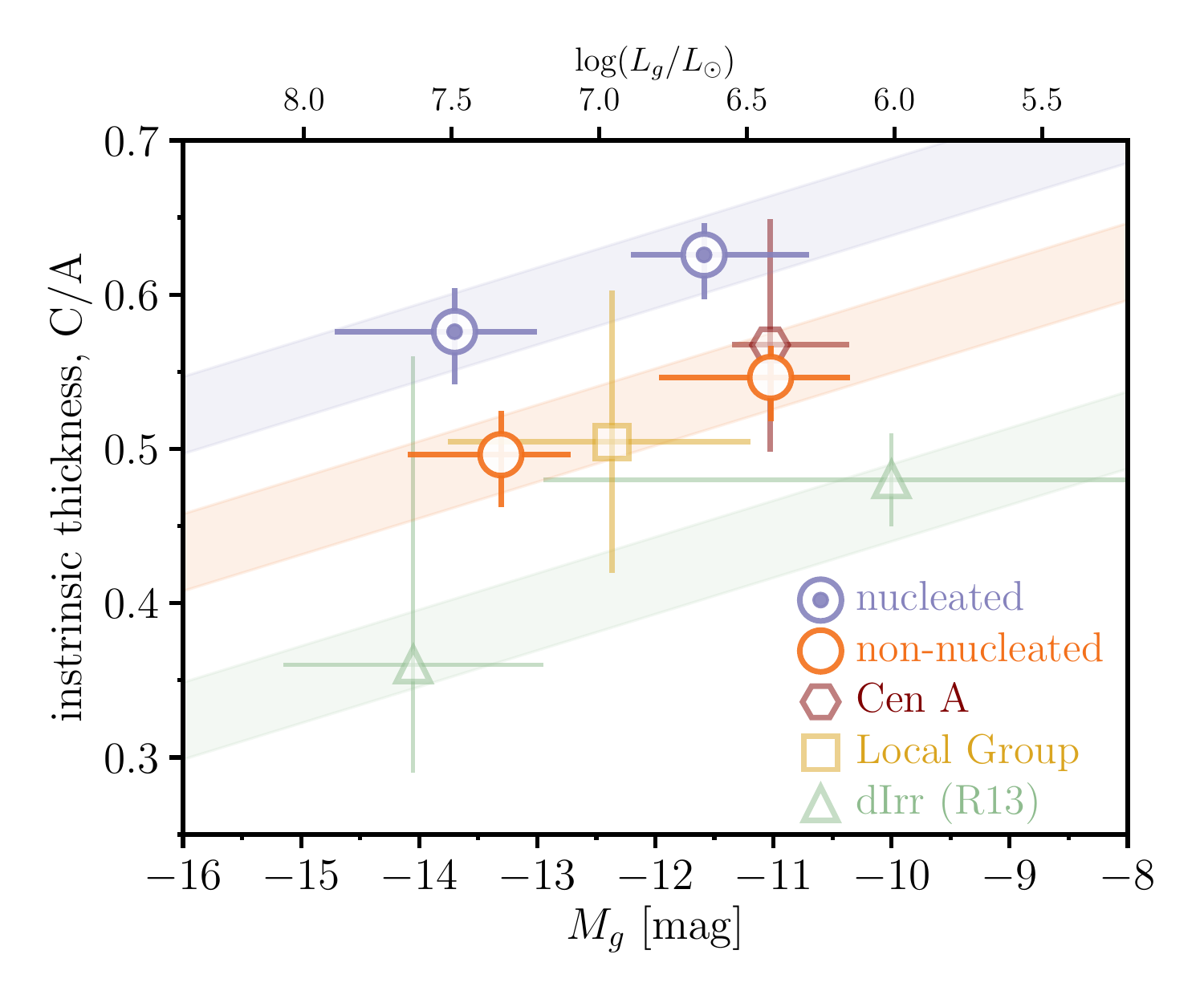}
\caption{Intrinsic thickness as a function of  absolute magnitude. Nucleated satellites are thicker at all luminosities compared to their non-nucleated counterparts, and fainter galaxies are systematically thicker as well, regardless of morphological type.
Shaded regions show that different types of dwarfs exhibit similar trends with luminosity.
}
\label{fig:ca}
\end{figure}

Fig.\,\ref{fig:ca} summarises the main finding of this study, showing the intrinsic thickness $C/A$ of all the subpopulations as a function of the median absolute magnitude of each subsample. 
X-axis error bars indicate the 16\% and 84\% percentiles of the magnitude distributions.
Non-nucleated dwarfs, in spite of being a much thicker population than regular spiral discs, are more flattened (at $\gtrsim$\,2$\sigma$) than nucleated systems at all luminosities.
This is consistent with early findings for brighter non-nucleated and nucleated Virgo dEs (\citealt{Ferguson1989,Ryden1994}; \citetalias{Lisker2007}), but here we show that the structural differences hold down to the faintest luminosities where NSCs still occur.
We also find that both nucleated and and non-nucleated dwarfs are thicker as they become fainter. 
The trend with luminosity was already found in previous work on faint field and LG galaxies \citep{Sanchez-Janssen2010a}, as illustrated in Fig.\,\ref{fig:ca} with the inclusion of flattening measurements for faint dIrrs by \citet[][triangles]{Roychowdhury2013}.
It is remarkable that the same behaviour holds in cluster environments, and separately for the nucleated and non-nucleated populations. 

The dIrrs are the flattest subpopulation among the dwarfs, and this speaks to a structural difference between satellites and dwarfs that inhabit low environmental densities.
However, the environmental dependence vanishes when we focus on satellites only and divide by morphological type (as traced by the presence of NSCs).
This is demonstrated in Fig.\,\ref{fig:ca} by the inclusion  of the CenA (hexagon) and LG (square) satellites. 
Despite the small sample sizes, these (non-nucleated) group satellites have intrinsic flattenings fully consistent with the subsamples of  non-nucleated cluster dwarfs at all luminosities.
It seems that at these low masses nucleation has a larger impact on the intrinsic shapes of group and cluster satellites than the environment.

\section{Implications and conclusions}
The main finding in this study is that the intrinsic flattening of faint ($10^{6} \lesssim L_{g}/L_\odot \lesssim 10^{8}$), quiescent satellites depends on galaxy luminosity {\it and} on nucleation, with the latter trend being slightly more significant than the former.
Once these two parameters have been accounted for, we find no evidence that the environment the galaxies live in, be it Virgo, Fornax, CenA or the LG, plays a relevant role in setting their 3D structure. 
This suggests that either their stellar distributions are mainly shaped by internal mechanisms, or that becoming a satellite of a more massive halo is a sufficient requirement, with specific halo-dependent environmental effects having little influence on their shapes. 

There is mounting evidence that faint galaxies form as thick, puffy systems \citep{Wheeler2015}.
Star formation in low-mass haloes is expected to occur in episodic bursts at almost all redshifts  \citep[][]{Muratov2015}, and the associated feedback-driven outflows pressurise gas and cause heating of the stellar orbits \citep{Pontzen2012,El-Badry2015}.
The progressively lower binding energies of fainter galaxies naturally allow for more impactful feedback effects and as a consequence the stellar distributions may become less flattened. 
The different intrinsic thickness of nucleated and non-nucleated cluster galaxies can also be explained by internal processes. 
The presence of compact stellar systems like NSCs and GCs is a clear indication that the host galaxies were actively forming stars at early cosmic times, when star formation rates (SFRs) and SFR surface densities were highest \citep{Peng2008,Kruijssen2015,Mistani2016}. 
The notion that nucleated galaxies are a biased subpopulation is further supported by the observed strong clustering and old ages of the stellar populations of intermediate-mass nucleated dEs in Virgo \citep{Ferguson1989,Lisker2008,Lisker2009}. 
As discussed above, this is accompanied by thicker intrinsic shapes compared to non-nucleated dEs, in the same way as found here for cluster satellites that are hundreds of times less luminous.

We therefore propose that the thicker shapes of nucleated faint galaxies in Virgo and Fornax are the result of preferential early and rapid formation. 
This is almost a necessity for objects that inhabit the inner cluster regions, for which typical infall times are of the order of several billion years.
But in this scenario the nucleated objects would be the most biased subpopulation, truly representing the first generation of cluster satellites.
A clear prediction from this hypothesis is that nucleated faint satellites should show evidence for enhanced abundance ratios compared to non-nucleated objects, which would be indicative of a formation history dominated by early intense starbursts \citep[][]{Liu2016}.
Early formation not only is characterised by explosive feedback events that result in a higher degree of dispersion support and thicker stellar distributions, but these early infallers have also spent a substantial amount of time in the central cluster environment, where more frequent and intense tidal interactions can further heat the stellar orbits \citep{Kazantzidis2011,Smith2015}.
While these mechanisms are arguably at place, it is important to note that some of the difference in flattening between satellites and dIrrs is almost certainly caused by the sustained stellar mass growth of the field dwarfs. 
The deepening of the potential together with the reduced gas accretion rates at more recent epochs result in weaker stellar feedback effects \citep{Muratov2015}, and this favours the development of stable and calm gas discs that are self-regulated by continuous star formation.





\bibliographystyle{mnras}
\bibliography{library} 

\begin{thebibliography}{}
\makeatletter
\relax
\def\mn@urlcharsother{\let\do\@makeother \do\$\do\&\do\#\do\^\do\_\do\%\do\~}
\def\mn@doi{\begingroup\mn@urlcharsother \@ifnextchar [ {\mn@doi@}
  {\mn@doi@[]}}
\def\mn@doi@[#1]#2{\def\@tempa{#1}\ifx\@tempa\@empty \href
  {http://dx.doi.org/#2} {doi:#2}\else \href {http://dx.doi.org/#2} {#1}\fi
  \endgroup}
\def\mn@eprint#1#2{\mn@eprint@#1:#2::\@nil}
\def\mn@eprint@arXiv#1{\href {http://arxiv.org/abs/#1} {{\tt arXiv:#1}}}
\def\mn@eprint@dblp#1{\href {http://dblp.uni-trier.de/rec/bibtex/#1.xml}
  {dblp:#1}}
\def\mn@eprint@#1:#2:#3:#4\@nil{\def\@tempa {#1}\def\@tempb {#2}\def\@tempc
  {#3}\ifx \@tempc \@empty \let \@tempc \@tempb \let \@tempb \@tempa \fi \ifx
  \@tempb \@empty \def\@tempb {arXiv}\fi \@ifundefined
  {mn@eprint@\@tempb}{\@tempb:\@tempc}{\expandafter \expandafter \csname
  mn@eprint@\@tempb\endcsname \expandafter{\@tempc}}}

\bibitem[\protect\citeauthoryear{Binggeli}{Binggeli}{1980}]{Binggeli1980}
Binggeli B.,  1980, \aap, 82, 289

\bibitem[\protect\citeauthoryear{Clauwens, Schaye, Franx  \& Bower}{Clauwens
  et~al.}{2017}]{Clauwens2017}
Clauwens B.,  Schaye J.,  Franx M.,   Bower R.~G.,  2017, \mn@doi [\mnras]
  {10.1093/mnras/sty1229}, 000, 1

\bibitem[\protect\citeauthoryear{C{\^{o}}t{\'{e}} et~al.,}{C{\^{o}}t{\'{e}}
  et~al.}{2006}]{Cote2006}
C{\^{o}}t{\'{e}} P.,  et~al., 2006, \mn@doi [\apjs] {10.1086/504042}, 165, 57

\bibitem[\protect\citeauthoryear{Danieli, van Dokkum, Merritt, Abraham, Zhang,
  Karachentsev  \& Makarova}{Danieli et~al.}{2017}]{Danieli2017}
Danieli S.,  van Dokkum P.,  Merritt A.,  Abraham R.,  Zhang J.,  Karachentsev
  I.~D.,   Makarova L.~N.,  2017, \mn@doi [\apj] {10.3847/1538-4357/aa615b},
  837, 136

\bibitem[\protect\citeauthoryear{Eigenthaler et~al.,}{Eigenthaler
  et~al.}{2018}]{Eigenthaler2018}
Eigenthaler P.,  et~al., 2018, \mn@doi [\apj] {10.3847/1538-4357/aaab60}, 855,
  142

\bibitem[\protect\citeauthoryear{El-Badry, Wetzel, Geha, Hopkins, Kere{\v{s}},
  Chan  \& Faucher-Gigu{\`{e}}re}{El-Badry et~al.}{2016}]{El-Badry2015}
El-Badry K.,  Wetzel A.,  Geha M.,  Hopkins P.~F.,  Kere{\v{s}} D.,  Chan
  T.~K.,   Faucher-Gigu{\`{e}}re C.-A.,  2016, \mn@doi [\apj]
  {10.3847/0004-637X/820/2/131}, 820, 131

\bibitem[\protect\citeauthoryear{Ferguson \& Sandage}{Ferguson \&
  Sandage}{1989}]{Ferguson1989}
Ferguson H.,  Sandage A.,  1989, \mn@doi [\apjl] {10.1086/185577}, 346, L53

\bibitem[\protect\citeauthoryear{Ferrarese et~al.,}{Ferrarese
  et~al.}{2012}]{Ferrarese2012}
Ferrarese L.,  et~al., 2012, \mn@doi [\apjs] {10.1088/0067-0049/200/1/4}, 200,
  4

\bibitem[\protect\citeauthoryear{Fitts et~al.,}{Fitts et~al.}{2018}]{Fitts2018}
Fitts A.,  et~al., 2018, \mn@doi [\mnras] {10.1093/mnras/sty1488}, 479, 319

\bibitem[\protect\citeauthoryear{Foreman-Mackey, Hogg, Lang  \&
  Goodman}{Foreman-Mackey et~al.}{2013}]{ForemanMackey2013}
Foreman-Mackey D.,  Hogg D.,  Lang D.,   Goodman J.,  2013, \mn@doi [\pasp]
  {10.1086/670067}, 125, 306

\bibitem[\protect\citeauthoryear{Foster et~al.,}{Foster
  et~al.}{2017}]{Foster2017}
Foster C.,  et~al., 2017, \mn@doi [\mnras] {10.1093/mnras/stx1869}, 472, 966

\bibitem[\protect\citeauthoryear{Franx, Illingworth  \& de Zeeuw}{Franx
  et~al.}{1991}]{Franx1991}
Franx M.,  Illingworth G.,   de Zeeuw T.,  1991, \mn@doi [\apj]
  {10.1086/170769}, 383, 112

\bibitem[\protect\citeauthoryear{Governato, Brook, Mayer  \& al.}{Governato
  et~al.}{2010}]{Governato2010}
Governato F.,  Brook C.,  Mayer L.,   al. E.,  2010, \mn@doi [\nat]
  {10.1038/nature08640}, 463, 203

\bibitem[\protect\citeauthoryear{Hayward \& Hopkins}{Hayward \&
  Hopkins}{2017}]{Hayward2015}
Hayward C.~C.,  Hopkins P.~F.,  2017, \mn@doi [\mnras] {10.1093/mnras/stw2888},
  465, 1682

\bibitem[\protect\citeauthoryear{Holden, van~der Wel, Rix  \& Franx}{Holden
  et~al.}{2012}]{Holden2012}
Holden B.,  van~der Wel A.,  Rix H.-W.,   Franx M.,  2012, \mn@doi [\apj]
  {10.1088/0004-637X/749/2/96}, 749, 96

\bibitem[\protect\citeauthoryear{Kaufmann, Wheeler  \& Bullock}{Kaufmann
  et~al.}{2007}]{Kaufmann2007}
Kaufmann T.,  Wheeler C.,   Bullock J.~S.,  2007, \mn@doi [\mnras]
  {10.1111/j.1365-2966.2007.12436.x}, 382, 1187

\bibitem[\protect\citeauthoryear{Kazantzidis, {\L}okas, Callegari, Mayer  \&
  Moustakas}{Kazantzidis et~al.}{2011}]{Kazantzidis2011}
Kazantzidis S.,  {\L}okas E.,  Callegari S.,  Mayer L.,   Moustakas L.,  2011,
  \mn@doi [\apj] {10.1088/0004-637X/726/2/98}, 726, 98

\bibitem[\protect\citeauthoryear{Kazantzidis, Mayer, Callegari, Dotti  \&
  Moustakas}{Kazantzidis et~al.}{2017}]{Kazantzidis2017}
Kazantzidis S.,  Mayer L.,  Callegari S.,  Dotti M.,   Moustakas L.~A.,  2017,
  \mn@doi [\apj] {10.3847/2041-8213/aa5b8f}, 836, L13

\bibitem[\protect\citeauthoryear{Krajnovi{\'{c}}, Emsellem, den Brok, Marino,
  Schmidt, Steinmetz  \& Weilbacher}{Krajnovi{\'{c}}
  et~al.}{2018}]{Krajnovic2018}
Krajnovi{\'{c}} D.,  Emsellem E.,  den Brok M.,  Marino R.~A.,  Schmidt K.~B.,
  Steinmetz M.,   Weilbacher P.~M.,  2018, \mn@doi [\mnras]
  {10.1093/mnras/sty1031}, 477, 5327

\bibitem[\protect\citeauthoryear{Kruijssen}{Kruijssen}{2015}]{Kruijssen2015}
Kruijssen J. M.~D.,  2015, \mn@doi [\mnras] {10.1093/mnras/stv2026}, 454, 1658

\bibitem[\protect\citeauthoryear{Li, Mao, Cappellari, Graham, Emsellem  \&
  Long}{Li et~al.}{2018}]{Li2018}
Li H.,  Mao S.,  Cappellari M.,  Graham M.~T.,  Emsellem E.,   Long R.~J.,
  2018, \mn@doi [\apj] {10.3847/2041-8213/aad54b}, 863, L19

\bibitem[\protect\citeauthoryear{Lisker, Grebel, Binggeli  \& Glatt}{Lisker
  et~al.}{2007}]{Lisker2007}
Lisker T.,  Grebel E.~K.,  Binggeli B.,   Glatt K.,  2007, \mn@doi [\apj]
  {10.1086/513090}, 660, 1186

\bibitem[\protect\citeauthoryear{Lisker, Grebel  \& Binggeli}{Lisker
  et~al.}{2008}]{Lisker2008}
Lisker T.,  Grebel E.,   Binggeli B.,  2008, \mn@doi [\aj]
  {10.1088/0004-6256/135/1/380}, 135, 380

\bibitem[\protect\citeauthoryear{Lisker et~al.,}{Lisker
  et~al.}{2009}]{Lisker2009}
Lisker T.,  et~al., 2009, \mn@doi [\apj] {10.1088/0004-637X/706/1/L124}, 706,
  L124

\bibitem[\protect\citeauthoryear{Liu et~al.,}{Liu et~al.}{2016}]{Liu2016}
Liu Y.,  et~al., 2016, \mn@doi [\apj] {10.3847/0004-637X/818/2/179}, 818, 179

\bibitem[\protect\citeauthoryear{{\L}okas, Majewski, Kazantzidis, Mayer,
  Carlin, Nidever  \& Moustakas}{{\L}okas et~al.}{2012}]{Lokas2012}
{\L}okas E.,  Majewski S.,  Kazantzidis S.,  Mayer L.,  Carlin J.,  Nidever D.,
    Moustakas L.,  2012, \mn@doi [\apj] {10.1088/0004-637X/751/1/61}, 751, 61

\bibitem[\protect\citeauthoryear{Ma et~al.,}{Ma et~al.}{2017}]{Ma2017}
Ma X.,  et~al., 2017, \mn@doi [\mnras] {10.1093/mnras/stx273}, 467, 2490

\bibitem[\protect\citeauthoryear{Mastropietro, Moore, Mayer, Debattista,
  Piffaretti  \& Stadel}{Mastropietro et~al.}{2005}]{Mastropietro2005}
Mastropietro C.,  Moore B.,  Mayer L.,  Debattista V.,  Piffaretti R.,   Stadel
  J.,  2005, \mn@doi [\mnras] {10.1111/j.1365-2966.2005.09579.x}, 364, 607

\bibitem[\protect\citeauthoryear{McConnachie}{McConnachie}{2012}]{McConnachie2012}
McConnachie A.~W.,  2012, \mn@doi [\aj] {10.1088/0004-6256/144/1/4}, 144, 4

\bibitem[\protect\citeauthoryear{Mistani et~al.,}{Mistani
  et~al.}{2015}]{Mistani2016}
Mistani P.~A.,  et~al., 2015, \mn@doi [\mnras] {10.1093/mnras/stv2435}, 455,
  2323

\bibitem[\protect\citeauthoryear{Muratov, Kere{\v{s}}, Faucher-Gigu{\`{e}}re,
  Hopkins, Quataert  \& Murray}{Muratov et~al.}{2015}]{Muratov2015}
Muratov A.~L.,  Kere{\v{s}} D.,  Faucher-Gigu{\`{e}}re C.-A.,  Hopkins P.~F.,
  Quataert E.,   Murray N.,  2015, \mn@doi [\mnras] {10.1093/mnras/stv2126},
  454, 2691

\bibitem[\protect\citeauthoryear{Navarro et~al.,}{Navarro
  et~al.}{2017}]{Navarro2017}
Navarro J.~F.,  et~al., 2017, \mn@doi [\mnras] {10.1093/mnras/sty497}, 000, 1

\bibitem[\protect\citeauthoryear{Ordenes-Brice{\~{n}}o
  et~al.,}{Ordenes-Brice{\~{n}}o et~al.}{2018}]{Ordenes-Briceno2018b}
Ordenes-Brice{\~{n}}o Y.,  et~al., 2018, \mn@doi [\apj]
  {10.3847/1538-4357/aac1b8}, 860, 4

\bibitem[\protect\citeauthoryear{Oser, Ostriker, Naab, Johansson  \&
  Burkert}{Oser et~al.}{2010}]{Oser2010}
Oser L.,  Ostriker J.~P.,  Naab T.,  Johansson P.~H.,   Burkert A.,  2010,
  \mn@doi [\apj] {10.1088/0004-637X/725/2/2312}, 725, 2312

\bibitem[\protect\citeauthoryear{Paudel, Lisker  \& Kuntschner}{Paudel
  et~al.}{2011}]{Paudel2011}
Paudel S.,  Lisker T.,   Kuntschner H.,  2011, \mn@doi [\mnras]
  {10.1111/j.1365-2966.2011.18256.x}, 413, 1764

\bibitem[\protect\citeauthoryear{Peng, Ho, Impey  \& Rix}{Peng
  et~al.}{2002}]{Peng2002}
Peng C.,  Ho L.,  Impey C.,   Rix H.-W.,  2002, \mn@doi [\aj] {10.1086/340952},
  124, 266

\bibitem[\protect\citeauthoryear{Peng et~al.,}{Peng et~al.}{2008}]{Peng2008}
Peng E.~W.,  et~al., 2008, \mn@doi [\apj] {10.1086/587951}, 681, 197

\bibitem[\protect\citeauthoryear{Pontzen \& Governato}{Pontzen \&
  Governato}{2012}]{Pontzen2012}
Pontzen A.,  Governato F.,  2012, \mn@doi [\mnras]
  {10.1111/j.1365-2966.2012.20571.x}, 421, 3464

\bibitem[\protect\citeauthoryear{Roediger et~al.,}{Roediger
  et~al.}{2017}]{Roediger2016}
Roediger J.~C.,  et~al., 2017, \mn@doi [\apj] {10.3847/1538-4357/836/1/120},
  836, 120

\bibitem[\protect\citeauthoryear{Roychowdhury, Chengalur, Karachentsev  \&
  Kaisina}{Roychowdhury et~al.}{2013}]{Roychowdhury2013}
Roychowdhury S.,  Chengalur J.,  Karachentsev I.,   Kaisina E.,  2013, \mn@doi
  [\mnras] {10.1093/mnrasl/slt123}, 436, L104

\bibitem[\protect\citeauthoryear{Ryden \& Terndrup}{Ryden \&
  Terndrup}{1994}]{Ryden1994}
Ryden B.,  Terndrup D.,  1994, \mn@doi [\apj] {10.1086/173960}, 425, 43

\bibitem[\protect\citeauthoryear{Salomon, Ibata, Martin  \& Famaey}{Salomon
  et~al.}{2015}]{Salomon2015}
Salomon J.-B.,  Ibata R.,  Martin N.,   Famaey B.,  2015, \mn@doi [\mnras]
  {10.1093/mnras/stv741}, 450, 1409

\bibitem[\protect\citeauthoryear{S{\'{a}}nchez-Janssen, M{\'{e}}ndez-Abreu  \&
  Aguerri}{S{\'{a}}nchez-Janssen et~al.}{2010}]{Sanchez-Janssen2010a}
S{\'{a}}nchez-Janssen R.,  M{\'{e}}ndez-Abreu J.,   Aguerri J. A.~L.,  2010,
  \mn@doi [\mnras] {10.1111/j.1745-3933.2010.00883.x}, 406, no

\bibitem[\protect\citeauthoryear{S{\'{a}}nchez-Janssen
  et~al.,}{S{\'{a}}nchez-Janssen et~al.}{2016}]{Sanchez-Janssen2016a}
S{\'{a}}nchez-Janssen R.,  et~al., 2016, \mn@doi [\apj]
  {10.3847/0004-637X/820/1/69}, 820, 69

\bibitem[\protect\citeauthoryear{S{\'{a}}nchez-Janssen
  et~al.,}{S{\'{a}}nchez-Janssen et~al.}{2018}]{Sanchez-Janssen2018}
S{\'{a}}nchez-Janssen R.,  et~al., 2018, arxiv:1812.01019

\bibitem[\protect\citeauthoryear{Sandage, Freeman  \& Stokes}{Sandage
  et~al.}{1970}]{Sandage1970}
Sandage A.,  Freeman K.,   Stokes N.,  1970, \mn@doi [\apj] {10.1086/150475},
  160, 831

\bibitem[\protect\citeauthoryear{Sanders \& Evans}{Sanders \&
  Evans}{2017}]{Sanders2017}
Sanders J.~L.,  Evans N.~W.,  2017, \mn@doi [\mnras] {10.1093/mnras/stx2116},
  472, 2670

\bibitem[\protect\citeauthoryear{Schombert}{Schombert}{2006}]{Schombert2006}
Schombert J.,  2006, \mn@doi [\aj] {10.1086/497964}, 131, 296

\bibitem[\protect\citeauthoryear{S{\'{e}}rsic}{S{\'{e}}rsic}{1968}]{Sersic1968}
S{\'{e}}rsic J.,  1968, {Atlas de galaxias australes}.
Cordoba, Argentina: Observatorio Astronomico, 1968

\bibitem[\protect\citeauthoryear{Simons et~al.,}{Simons
  et~al.}{2017}]{Simons2017}
Simons R.~C.,  et~al., 2017, \mn@doi [\apj] {10.3847/1538-4357/aa740c}, 843, 46

\bibitem[\protect\citeauthoryear{Smith, Davies  \& Nelson}{Smith
  et~al.}{2010}]{Smith2010}
Smith R.,  Davies J.,   Nelson A.,  2010, \mn@doi [\mnras]
  {10.1111/j.1365-2966.2010.16545.x}, 405, 1723

\bibitem[\protect\citeauthoryear{Smith et~al.,}{Smith et~al.}{2015}]{Smith2015}
Smith R.,  et~al., 2015, \mn@doi [\mnras] {10.1093/mnras/stv2082}, 454, 2502

\bibitem[\protect\citeauthoryear{Spengler et~al.,}{Spengler
  et~al.}{2017}]{Spengler2017}
Spengler C.,  et~al., 2017, \mn@doi [\apj] {10.3847/1538-4357/aa8a78}, 849, 55

\bibitem[\protect\citeauthoryear{Taylor et~al.,}{Taylor
  et~al.}{2017}]{Taylor2017}
Taylor M.~A.,  et~al., 2017, \mn@doi [\mnras] {10.1093/mnras/stx1021}, 469,
  3444

\bibitem[\protect\citeauthoryear{Taylor et~al.,}{Taylor
  et~al.}{2018}]{Taylor2018}
Taylor M.~A.,  et~al., 2018, \mn@doi [\apj] {10.3847/2041-8213/aae88d}, 867,
  L15

\bibitem[\protect\citeauthoryear{Turner, C{\^{o}}t{\'{e}}, Ferrarese,
  Jord{\'{a}}n, Blakeslee, Mei, Peng  \& West}{Turner
  et~al.}{2012}]{Turner2012}
Turner M.~L.,  C{\^{o}}t{\'{e}} P.,  Ferrarese L.,  Jord{\'{a}}n A.,  Blakeslee
  J.~P.,  Mei S.,  Peng E.~W.,   West M.~J.,  2012, \mn@doi [\apjs]
  {10.1088/0067-0049/203/1/5}, 203, 5

\bibitem[\protect\citeauthoryear{Venhola et~al.,}{Venhola
  et~al.}{2018}]{Venhola2018}
Venhola A.,  et~al., 2018, pp 1--31

\bibitem[\protect\citeauthoryear{Wheeler et~al.,}{Wheeler
  et~al.}{2017}]{Wheeler2015}
Wheeler C.,  et~al., 2017, \mn@doi [\mnras] {10.1093/mnras/stw2583}, 465, 2420

\makeatother
\end{thebibliography}





\bsp	
\label{lastpage}
\end{document}